\documentclass[
superscriptaddress,twocolumn,
 amsmath,amssymb,
 aps,prl,
]{revtex4-2}

\usepackage{graphicx}
\usepackage{dcolumn}
\usepackage{bm}

\usepackage[dvipsnames]{xcolor}
\usepackage{textcomp}
\usepackage{gensymb}

\usepackage[colorlinks=true,
            linkcolor=blue, 
            citecolor=blue, 
            urlcolor=blue, 
            breaklinks=true]{hyperref}

\newcommand{\dipc}{Donostia International Physics Center (DIPC), E-20018 Donostia-San Sebasti\'an, Spain}
\newcommand{\ikerbasque}{IKERBASQUE, Basque Foundation for Science, E-48013 Bilbao, Spain}
\newcommand{\dtu}{Department of Physics, Technical University of Denmark, DK-2800 Kgs.~Lyngby, Denmark}

\newcommand{\etal}{\mbox{\textit{et al.}}~}

\newcommand{\ie}{{\em i.\,e. }}

\begin{document}

\title{Twisted nanoporous graphene/graphene bilayers:\\electronic decoupling and chiral currents}

\author{Xabier Diaz de Cerio}
\affiliation{\dipc}
\author{Aleksander Bach Lorentzen}
\affiliation{\dtu}
\author{Mads Brandbyge}
\affiliation{\dtu}
\author{Aran Garcia-Lekue}
\affiliation{\dipc}
\affiliation{\ikerbasque}

\date{\today}

\begin{abstract}

We investigate bilayers of nanoporous graphene (NPG), laterally bonded carbon nanoribbons, and graphene. The electronic and transport properties are explored as a function of the interlayer twist angle using an atomistic tight-binding model combined with non-equilibrium Green's functions. At  small twist angles  ($\theta \lesssim 10^\circ$), NPG and graphene are strongly coupled, as revealed by the hybridization of their electronic bands. As a result, when electrons are point-injected in NPG, a substantial interlayer transmission occurs and an electronic Talbot-like interference pattern appears in the current flow on both layers. Besides, the twist-induced mirror-symmetry-breaking leads to chiral features in the injected current. Upon increasing the twist angle, the coupling is weakened and the monolayer electronic properties are restored. Furthermore,  we demonstrate the emergence of resonant peaks in the electronic density of states for small twist angles, allowing to probe the twist-dependent interlayer coupling via scanning tunneling microscopy.
 
\end{abstract}

\maketitle

Quantum confinement effects in nanostructured graphene are responsible
for a wide range of physical phenomena, such as the opening of band gaps and the emergence of topological states and magnetism \cite{Son2006,Yang2007,Nakada1996,Yazyev2011,Cao2017}, 
both of fundamental and practical relevance in nanoelectronics or quantum spintronics.
In particular, quasi-1D
graphene nanoribbons (GNRs) can exhibit semiconducting character while preserving 
ballistic and coherent electronic transport properties \cite{Berger2006,Li2008,Baringhaus2014}, rendering them ideal to realize
fast and efficient 
devices. GNRs can be 
fused laterally to form nanoporous graphene (NPG) \cite{Moreno2018},
a 2D array of covalently bonded identical and parallel nanoribbons. Such NPG superlattices can be finely tuned  by modifications
in the structural or chemical
conformation of 
their parent GNRs\cite{Jacobse2020,Piquero2024, Tenorio2022}.

Remarkably, 
NPG  shows  
a strong
in-plane 
anisotropy for electron states near the valence and conduction band, 
the inter-ribbon coupling being weaker than the intra-ribbon kinetic 
energy\cite{Moreno2018}. Due to its
anisotropic electronic structure,
electrons propagating in NPG exhibit the so-called electronic Talbot interference effect, known from light propagation in coupled waveguides\cite{Calogero2019a}. 
This arises from the  interference between Bloch states of same energy but different longitudinal (i.e. along the nanoribbon) 
wave-vector, and
strongly depends on the 
inter-connections between 
GNRs \cite{Calogero2019b,Alcon2021,Moreno2023}.

In order to fully exploit its potential, NPG should
ideally be on a substrate which preserves or enhances its unique, 
anisotropic electron transport behaviour.
Motivated by the wide range of intriguing phenomena recently discovered in bilayer graphene \cite{Cao2018a,Cao2018b},
stacking NPG on graphene emerges
as a particularly promising
avenue of research. Recent theoretical studies have analyzed some specific properties of aligned NPG/graphene bilayers \cite{Antidormi2021,Lee2022}. 
Antidormi \etal\cite{Antidormi2021}
find that the proximity of NPG induces a renormalization of the graphene Dirac cone leading to anisotropic optical and electrical conductivities at low energies. In analogy to bilayer graphene
 \cite{Zhang2009}, Lee \etal\cite{Lee2022} predict
   a band-gap opening at the Dirac point by applying an electric field perpendicular to the bilayer.
However, these previous works considered aligned NPG/graphene bilayers with an AA or AB relative stacking and, 
thus, the impact of twisting the layers remains unaddressed. 

\begin{figure*}[th]
\includegraphics[width=0.9\linewidth]{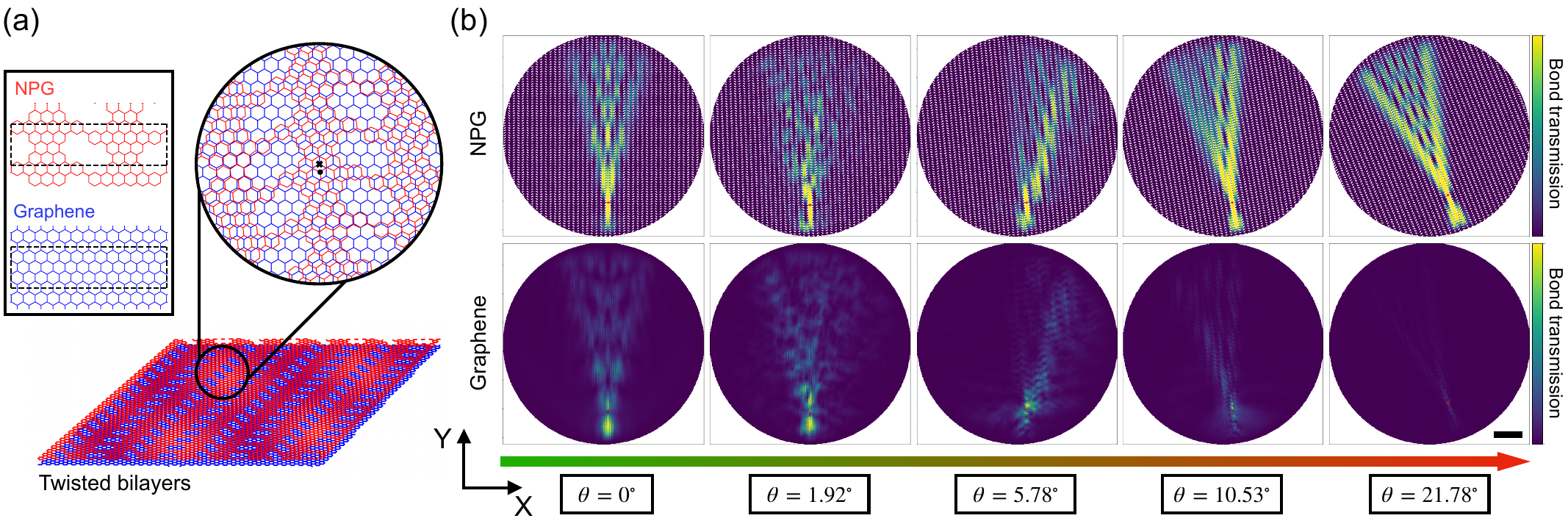}
\caption{\label{Fig1}System set-up and bond-transmissions. (a) Atomic structure of NPG (red), graphene (blue) and a twisted NPG/graphene bilayer. The black cross and the black point indicate the rotation axis and injection site, respectively. (b) Real-space bond-transmission maps in twisted NPG/graphene bilayers at an energy $E = -0.4$ eV. Top row: NPG layer. Bottom row: graphene layer. The red dot indicates the injection point in the NPG layer. Scale bar is 10nm.} 
\end{figure*}

In this Letter, we explore the tunability of the 
 electronic and transport properties of 
 NPG/graphene bilayers as a function of the interlayer twist angle. Based on
 calculations combining an atomistic tight-binding model and Non-equilibrium Green's Functions (NEGF), we show
 a progressive reduction of the electronic coupling upon 
 increasing the twist. 
 This is reflected in the modulation of
 the intra- and 
 inter-layer electronic currents when electrons are injected in the bilayer by point contacting NPG.
 For twist angles below $\sim 10$\degree, 
the twist-induced mirror-symmetry-breaking leads to a chiral flow pattern in the Talbot-like, point-injected current 
propagation along the NPG ribbons, which is indeed imprinted in the underlying graphene layer.
Importantly, for larger twist angles, NPG and graphene are found to be effectively decoupled and exhibit monolayer-like behaviour.

The bilayer system studied in this work
is shown in Fig. \ref{Fig1}a. It is 
composed of a NPG 
structure, as the one synthesized with atomic precision  
in Ref. \cite{Moreno2018}, stacked on top of 
a single graphene layer.
Initially, we consider both monolayers 
to be aligned in a commensurate AB or Bernal stacking sequence,
corresponding to a single unit-cell of NPG (black box in
Fig. \ref{Fig1}a). 
Next, we allow for rotations at commensurate angles 
$\theta$ of the NPG around a
vertical axis 
located 
at the position marked by
a cross in Fig. \ref{Fig1}a, while keeping
the graphene layer fixed.
For all twist angles, 
a C-C distance of $a=1.42$\,Å for neighbouring atoms within each layer and
an interlayer separation of $d=3.35$\,Å are employed (see Supplemental Material (SM)  for further details on the atomic set-up \cite{SM}).

The electronic structure of the system 
is described using a $p_z$-orbital tight-binding Hamiltonian with intralayer hoppings restricted to first nearest neighbours. Interlayer terms are 
computed through Slater-Koster-type two-center
bond-integrals
in which hopping amplitudes depend only on the planar projection of the interatomic distance 
\cite{Slater1954}. This
parametrization of the interlayer coupling terms has been successfully applied to characterize the band structure and  Fermi velocity renormalization in twisted bilayer graphene \cite{Lopes2007,Trambly2010}, as well
as to describe electron transport in crossed graphene nanoribbons \cite{Sanz2020, sanz2022}. All parameters were fitted to reproduce the band structure of NPG as obtained using Density Functional Theory (DFT). 
Based on this tight-binding model Hamiltonian, 
electronic transport simulations are carried out using the NEGF formalism. 
This approach allows us to  
treat highly realistic experimental set-ups containing large
NPG/graphene systems ($\approx$\,100 nm in diameter)\cite{Calogero2018}. 
We refer to the SM for details of the
DFT calculations, the tight-binding model, and transport simulations\cite{SM}. 

The effect of the interlayer twisting on electron transport can 
be directly visualized by computing the 
bond transmissions; \ie the electron transmission between the $p_z$-orbitals
of each carbon atom when  electrons are injected from a metallic tip in contact with NPG.
Figure \ref{Fig1}b shows in-plane bond transmissions in NPG and graphene layers 
for different twist angles. The tip injection point is located at the position of the red dot. Electrons are injected at an energy ($E=-0.4$ eV), lying within the range 
of anisotropic NPG bands, for which 
Talbot-interference patterns are known to 
emerge due to the inter-GNR coupling \cite{Calogero2019a}.
This characteristic "fingerprint" is reproduced 
for the untwisted case (AB stacking, $\theta=0^{\circ}$), although
significantly smeared out by its coupling to graphene. 
Remarkably, as a result of the interlayer coupling,
there is a significant vertical tranmission and
the Talbot-like pattern is clearly imprinted in the underlying graphene layer, giving rise to highly anisotropic current flow in an
otherwise isotropic system. 
Upon twisting NPG by
 $\theta=1.92^{\circ}$, both the Talbot pattern in NPG
and the imprinted current flow in graphene 
get further smeared out, indicating an enhanced interlayer coupling.

For larger twist angles, on the contrary, 
the interlayer transmission is gradually suppressed and, thus, 
the imprinted electron flow in graphene is also quenched.  This is accompanied 
by an increase in the electron transmission within the NPG layer. More interestingly, 
we observe strong asymmetries 
of the Talbot-like interference 
pattern with respect to the contacted GNR 
(see $\theta = 5.78^\circ$ and $10.53^\circ$), which would be reversed for $\theta \rightarrow -\theta$.
Such asymmetric features in the current can be
related to the breaking of in-plane mirror symmetry around the GNR of the injection
(see SM~\cite{SM}), introducing a chirality in the twisted bilayer.
Such asymmetries disappear and an
unperturbed Talbot-like pattern is recovered 
for a twist angle of $\theta = 21.78^\circ$, 
suggesting a complete decoupling
between NPG and graphene.

\begin{figure*}
\includegraphics[width=\linewidth]{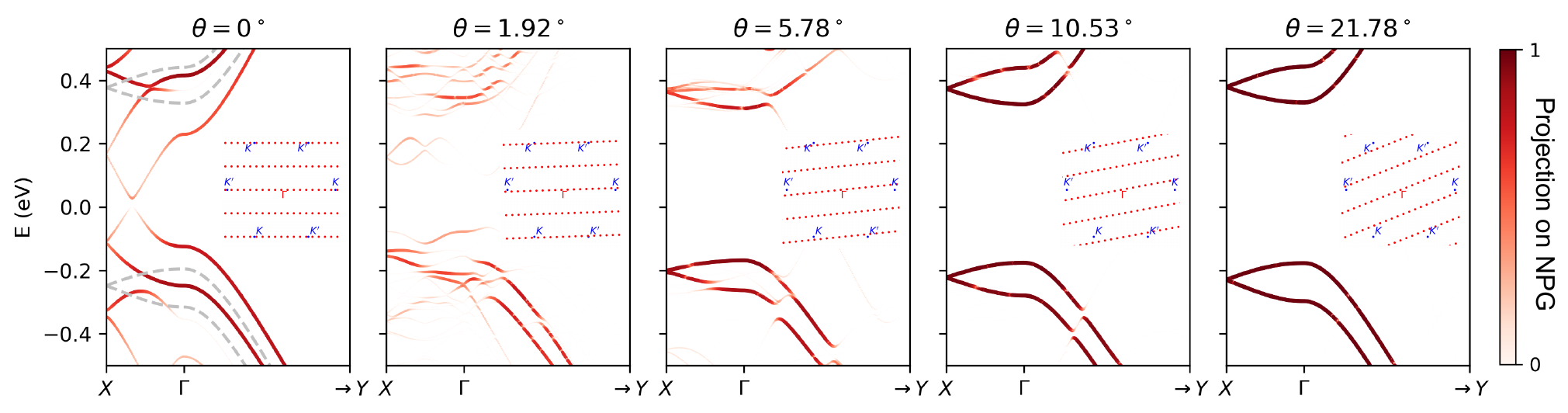}
\caption{\label{Fig2} Bandstructure of NPG/graphene at different $\theta$, projected on NPG and unfolded to its Brillouin zone. Unfolded bands are represented in directions across ($X \rightarrow \Gamma$) and along ($\Gamma \rightarrow Y$) the NPG ribbon axis. The band structure of monolayer NPG is plotted as a reference in the panel corresponding to $\theta = 0^\circ$ (gray dashed lines). Insets show the $K/K'$ points of graphene (blue) and extended $\Gamma$ points of NPG rotated by the corresponding angle (red) in a generalized momentum space before folding.}
\end{figure*}

To understand the electron transmission in 
more detail, we analyze 
the bandstructure of the NPG/graphene bilayer,
projected onto the electronic states of NPG and 
unfolded to its Brillouin zone.
Its evolution as a function of the twist angle is shown in Fig. \ref{Fig2}. 
For AB stacking ($\theta = 0^\circ$)
a strong  
hybridization between NPG and graphene is 
observed  upon folding of graphene $K$($K'$) points to $2/3 (\pi/L_x)$ in the $\Gamma X$ path, where $L_x = 13\sqrt{3}a$ is the NPG lattice constant in the direction accross the GNRs.
This is
reflected in the significant contribution of
the NPG states to the graphene-like bands around Fermi level.
Notably, despite their energy renormalization  at the 
$\Gamma$-point,
the NPG-like bands retain their anisotropic character 
along $\Gamma X$ and $\Gamma Y$ directions (see comparison with
bandstructure of monolayer NPG
in left panel of Fig. \ref{Fig2}). For a twist angle of $\theta=1.92^\circ$, 
the bandstructure of the system gets notably modified, with no appreciable graphene-like bands around Fermi level.
Besides,
NPG bands are strongly perturbed and reveal numerous avoided crossings and weakly dispersing states, which at negative energies emerge predominantly close to the top of the valence band ($\approx -0.1$\,eV). As $\theta$ increases further, avoided crossings 
appear at increasing energies and become weaker, leading to almost NPG single-layer-like dispersion at $\theta=21.78^\circ$. 
This evolution in the band structure, showing an effective decoupling between NPG and graphene upon increasing the twist angle, 
explains the gradual suppression of the interlayer transmission and the concomitant single-layer transport behaviour along NPG 
for large values of 
$\theta$ shown in Fig. \ref{Fig1}.

Within the tight-binding model used in this work, the interlayer hopping terms depend solely on the planar projection of the inter-atomic separation and, thus,  only single-layer NPG and graphene eigenstates with the same momentum can couple \cite{Bistritzer2010}. 
In particular, as anisotropic NPG bands are centered around $\Gamma$,
further insight into
their coupling
to the graphene Dirac bands
can be obtained by exploring the momentum separation
between periodically extended
$\Gamma$ points of NPG and
$K/K'$ points of graphene.
Besides, as the Fourier transform of the interlayer hopping parameter
 is expected to decay rapidly in the scale of $1/d$, only
 the first $K/K'$ points
 of graphene will be 
relevant \cite{Bistritzer2011}.
The insets in Fig. \ref{Fig2}
reveal the evolution 
of the pertinent momentum separations
as a function of the interlayer twist angle. 
For $\theta=0^\circ$, $K/K'$ points (blue dots) lie $2/3 (\pi/ L_x)$
away from the closest $\Gamma$ points of NPG (red dots) and 
the Dirac cones cross the anisotropic NPG bands around $E \approx -0.2$ eV, 
giving rise 
to significant hybridization at low energies.
Upon increasing $\theta$,
there is an initial decrease in the momentum separation
between $K/K'$ and 
$\Gamma$ points
(see inset for $\theta=1.92^\circ$).
For larger angles, instead,
the separation starts an overall increase
 that causes interlayer band crossings to be gradually shifted to higher energies beyond the range of interest. 
This trend is 
reserved 
at $\theta=21.78^\circ$, where the bands of both layers overlap again in momentum within the energy window of interest. 
At such large $\theta$, however, the real-space overlap integral between rotated NPG and unrotated graphene wave-functions is suppressed by symmetry and the coupling becomes negligible as well \cite{SM}.

We can obtain a quantitative description of 
the interlayer coupling by computing the so-called Inverse Participation
Ratio (IPR) which is a commmonly used measure of localization \cite{Bell1970,Edwards1972}.
For an eigenstate of the NPG/graphene 
bilayer, $\psi_\nu$, with band index and crystal momentum contained in $\nu$,
the IPR is given by $A_\nu = [P^2_{Gr}(\nu) + P^2_{NPG}(\nu)]^{-1}$, where $P_{l}(\nu)$ is the probability of finding the eigenstate $\psi_\nu$ on layer $l$. 
For an eigenstate that is a coherent combination of the two layers with equal probability ($P_{Gr}(\nu) = P_{NPG}(\nu)=1/2$), the IPR is maximized to $A_\nu = 2$. For full decoupling, the eigenstates will be localized in a single 
layer ($P_{Gr}(\nu) = 0$ or $1$ and $P_{NPG}(\nu)=1$ or $0$, respectively) and $A_\nu = 1$. 
The IPR of NPG/graphene at a given energy, $E$, is then provided by the weighted arithmetic mean $A(E) = \sum_\nu A_\nu \delta(E-E_\nu) / \sum_\nu \delta(E-E_\nu)$.

\begin{figure}
\includegraphics[width=\linewidth]{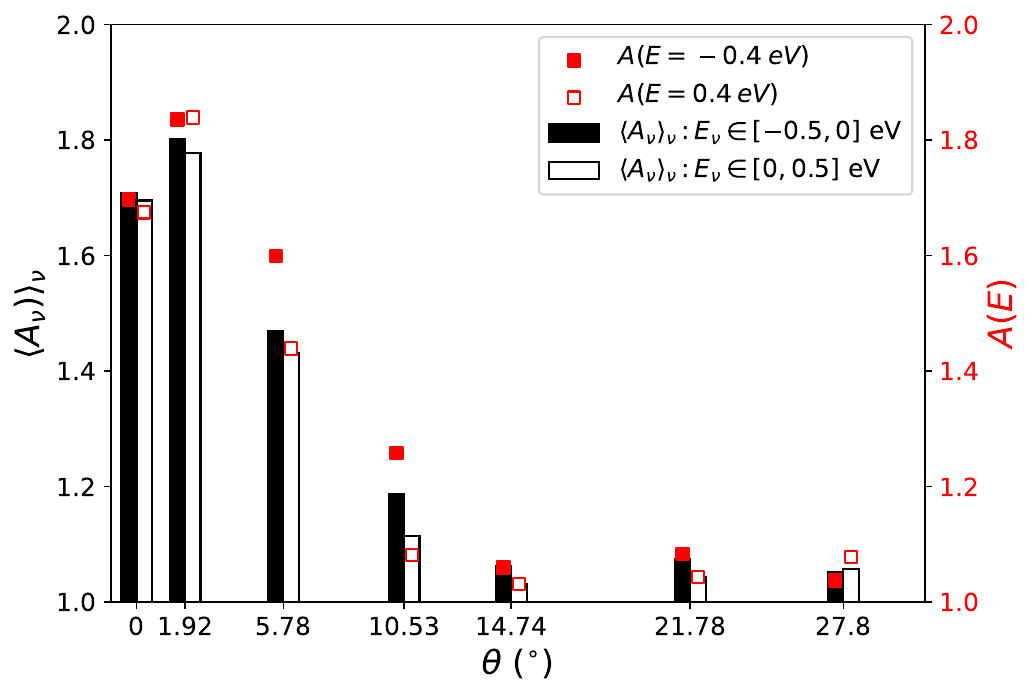}
\caption{\label{Fig3} Energy-resolved and averaged IPR of NPG/Graphene at different $\theta$. The average of $A_\nu$ is sepparately taken over states at negative energies $E_\nu \in [-0.5,0]$ eV (black solid bars) and positive energies $E_\nu \in [0.,0.5]$ eV (black empty bars). The energy-resolved IPR, $A(E)$, is evaluated at $E=-0.4$ eV (red solid squares) and $E=0.4$ eV (red empty squares).} 
\end{figure}

In Fig. \ref{Fig3} we represent
$A_\nu$ as a function of the twist angle for $E=-0.4$ eV and $+0.4$ eV, and averaged over
an energy window of 0.5\,eV
above and below the Fermi level. 
For $E=-0.4$ eV, the overall decrease of the IPR upon increasing $\theta$ 
is in agreement with the weakening of the interlayer coupling revealed in the analysis of the band 
structure (Fig.~\ref{Fig2}) as well as on the electron transmission (Fig.~\ref{Fig1}).
In fact, we observe the same behaviour for $E=0.4$ eV and
the energy-averaged IPR, suggesting that the twist-dependent decoupling is not tied to a specific energy. In all cases, the IPR shows an overall decrease of hybridization upon increasing the interlayer twist angle.
The trend is however non-monotonic, with a remarkable enhancement of the hybridization at $\theta=1.92^\circ$.
 This result is consistent with the
analysis of the momentum separation
 described above,
 which revealed 
 a larger
  overlap of relevant NPG and graphene electronic states
at such small
twist angle.

\begin{figure}
\includegraphics[width=\linewidth]{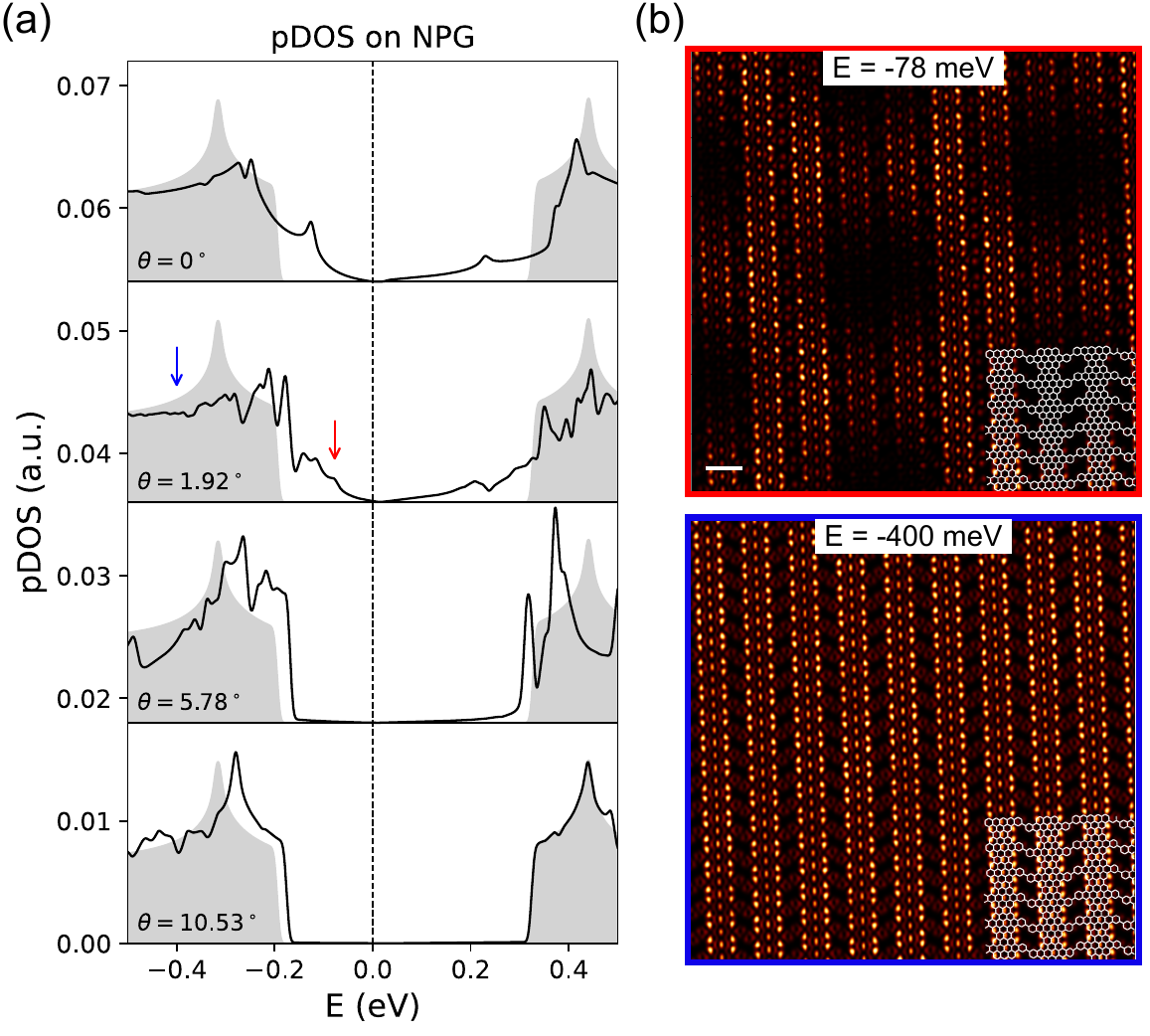}
\caption{\label{Fig4} (a) Density of states of NPG/Graphene projected on NPG at different $\theta$. The DOS of monolayer NPG is shown as a reference in all cases (gray shaded curve). (p)DOS is represented per atom. (b) Local DOS on NPG for $\theta=1.92^\circ$ evaluated at the energies indicated by the red ($E=-78$ meV) and blue arrows ($E=-400$ meV) in (a). The scale bar in the top panel is 1 nm and the NPG geometry is overlaid in the bottom right corner of both panels.}
\end{figure}

Spectroscopic signatures of interlayer NPG/graphene coupling can be obtained
from the 
density of states  
projected (pDOS) on NPG, as shown in Fig. \ref{Fig4}.
For $\theta=0^\circ$ and $1.92^\circ$, where 
a strong hybridization between NPG and graphene exists, the pDOS 
shows sharp peaks 
within the energy gap and around 
the band onsets of 
monolayer NPG (the DOS of NPG is highlighted with a
grey background in Fig.\ref{Fig4}),
in line with the remarkable energy renormalization of frontier NPG-like states shown in Fig. \ref{Fig2}.
 As $\theta$ increases further, these resonances 
 gradually move out of NPG energy gap 
 and the pDOS closely resembles the spectrum of monolayer NPG. 
 
 Additional information about the 
 spatial electronic distribution of the interlayer coupling induced resonant peaks
 can be extracted from the local density of states (LDOS) evaluated at selected energy positions.
Fig.\ref{Fig4}b displays the 
LDOS for $\theta=1.92^\circ$
at two representative energies,
$E=-400$ meV and  $E=-78$ meV,
outside and inside NPG energy gap, respectively
 (blue and red arrows
in Fig.\ref{Fig4}a).
For $E=-400$ meV, the LDOS looks very similar to that of the valence band of monolayer NPG (see Supplemental Material \cite{SM}).
For $E=-78$ meV, instead, 
the LDOS 
exhibits a long-range order modulation
that can be ascribed to the interaction of NPG with the underlying graphene layer.
In particular, the LDOS at this resonant energy
within the gap of NPG
vanishes around the domain walls between regions of local AB and BA stacking that are perpendicular to the 
axis of the nanoribbons.
Remarkably, the emergence of resonant peaks in the energy-resolved DOS and 
long-range order in the local  
distribution of selected electronic states, 
could be experimentally probed  
using dI/dV point 
spectroscopy and dI/dV mapping 
via scanning tunneling microscopy (STM) \cite{Brihuega2012,Wong2015}.

In summary, using
tight-binding and NEGF calculations, we have analyzed 
the effect of the interlayer twist angle 
on the electronic and transport properties
of NPG/graphene bilayers.
At small twist angles, 
NPG and graphene are found to be strongly coupled, 
as revealed by the hybridization of
their electronic bands. This results in a substantial
interlayer transmission, giving rise
to anisotropic, chiral point-injected current flow with signatures of Talbot-like interference on both layers.
Upon increasing the twist angle,
the interlayer coupling is weakened and NPG gradually recovers its monolayer electronic characteristics.  We quantify the behavior using IPR. Our findings suggest that experiments using STM or the Quantum Twisting Microscope (QTM) \cite{Inbar2023} may address the role of twist on interlayer coupling.

\begin{acknowledgments}
This work was supported by grant TED2021-132388B-C44 funded 
by MCIN/AEI/10.13039/501100011033 and Unión Europea Next
Generation EU/PRTR, and grant PID2022-140845OB-C66 funded by
MCIN/AEI/10.13039/501100011033 and FEDER Una manera de hacer
Europa (A.G.-L).A.G.-L also acknowledges the financial support received
from the IKUR Strategy under the collaboration agreement between
Ikerbasque Foundation and DIPC on behalf of the Department of Education
of the Basque Government. ABL, MB acknowledge funding from the Independent Research Fund Denmark (0135-00372A).
\end{acknowledgments}

\bibliography{bibliography}

\end{document}


\title{Supplemental Material: Twisted nanoporous graphene/graphene bilayers:\\ electronic decoupling and chiral currents
}
\author{Xabier Diaz de Cerio}
\affiliation{\dipc}
\author{Aleksander Bach Lorentzen}
\affiliation{\dtu}
\author{Mads Brandbyge}
\affiliation{\dtu}
\author{Aran Garcia-Lekue}
\affiliation{\dipc}
\affiliation{\ikerbasque}

\date{\today}

\maketitle

\tableofcontents

\section{Methods} 

Here we describe details of the atomic structure of the bilayer system, the Slater-Koster-based tight-binding model, the DFT calculations used to fit the model parameters, and the electronic transport simulations based on Non-Equilibrium Green's Functions.

\subsection{Atomic structure of NPG/graphene bilayers}

The bilayer system studied in this work is 
composed of a NPG 
structure, as the one synthesized with atomic precision  
in Ref. \cite{Moreno2018}, stacked on top of 
a single graphene layer. The in-plane C-C distance is $a=1.42$\,Å in both layers, and the interlayer separation is $d=3.35$\,Å, which are the values typically used to model twisted bilayer graphene \cite{Lopes2007,Trambly2010}.
Initially, we consider both monolayers 
to be aligned in an AB or Bernal stacking sequence for which commensuration is achieved for a single unit-cell of NPG. Next, we consider rotations at commensurate angles 
$\theta$ of the NPG layer around a
vertical axis crossing a graphene site located below the central hexagon of the 7-atom-wide section of a NPG ribbon (see Fig. 1a in the main text), while the graphene layer is kept fixed. The geometries of NPG/graphene bilayers under commensurate rotations were generated combining the SISL package \cite{zerothi_sisl} with the Supercell-core software \cite{Necio2020}. Atomic structure relaxations, which crucially affect the position of "magic-angles" in twisted bilayer graphene \cite{Carr2019}, are not considered in our model, and in-plane C-C distance and interlayer separations are kept fixed for all twist-angles.

\subsection{Density Functional Theory calculations}

Electronic structure calculations were performed using DFT as implemented in the SIESTA code \cite{Soler2002,Garcia2020} as a reference to fit the tight-binding model. In order to compare the results to the model, geometry and lattice relaxations were not consider, and the C-C distance and interlayer separation were fixed to $a=1.42$ Å and $d=3.35$ Å, respectively. Besides, a vacuum region of 30 Å was included in the direction perpendicular to the bilayers in order to avoid spurious effects between periodic images. Core electrons were addressed by norm-conserving Trouiller-Martins pseudopotentials \cite{Troullier1991}, while a linear combination of atomic orbitals was used for valence electrons. A double-$\zeta$ polarized basis set was employed, with the basis orbitals range defined by a 0.01 Ry energy shift \cite{Junquera2001}. Exchange-correlation was treated within the generalized gradient approximation by Perdew, Burke and Ernzerhof (GGA-PBE) \cite{Perdew1996}. The Brillouin zone was sampled by a 5x19x1 Monkhorst-Pack \textit{k}-grid \cite{Monkhorst1976} and the real-space grid was defined by a 400 Ry mesh-cutoff. 

\subsection{Tight-binding model and parameters}

\begin{figure*}[t]
\begin{center}
\includegraphics[width=2in]{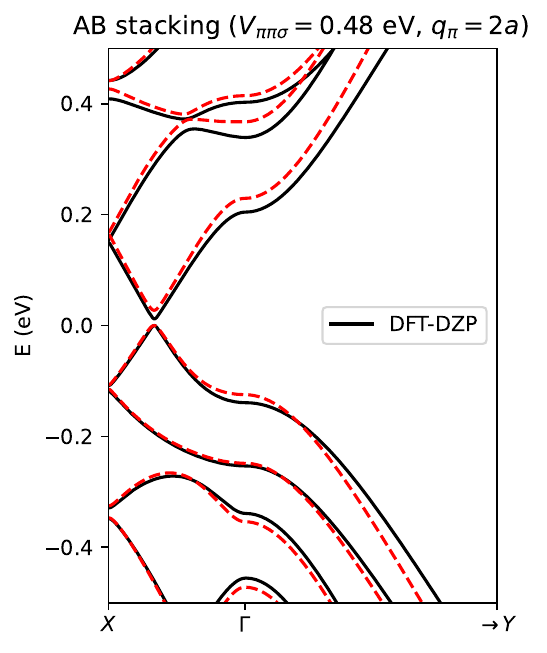}
\caption{Band structure comparison between DFT (black solid line) and the tight-binding model used in this work (red dashed line) for an aligned NPG/graphene in the AB stacking configuration.
}
\label{FigS1}
\end{center}
\end{figure*}

We describe the electronic structure of NPG/graphene bilayers with commensurate twist-angles $\theta$ using a $p_z$-orbital tight-binding Hamiltonian 
\begin{equation}\label{Eq1}
    H = \sum_{i}{\epsilon_{i} c_{i}^\dagger c_i} + \sum_{ij}{t_{ij} (c_i^\dagger c_j + \text{H.c.})},
\end{equation}
where $c_i$ ($c_i^\dagger$) is the annihiliation (creation) operator on site $i$, $\epsilon_i$ is the on-site energy of a $p_z$-orbital at site $i$, and $t_{ij}$ is the hopping amplitude between orbitals in sites $i$ and $j$.

Intralayer hoppings are restricted to first nearest neighbours (1NN) such that $t_{ij}=t_{\parallel}$, where $t_{\parallel} = -2.7$ eV. This approximation provides a good description of the Dirac cones of monolayer graphene, as well as the semiconducting and anisotropic band structure of monolayer NPG in the energy region of interest for this work ($|E| \leq 0.5$ eV). On-site energies have been set to zero in all orbitals except
at the pore edges of NPG,
where a positive $\epsilon_i = 0.2$ eV 
shift accounting for
the electrostatic potential increase upon C-H bond formation
has been applied.
Moreover, this latter assignment
reproduces the
relative energy misalignment between the graphene Dirac point  
and the middle of the NPG gap as provided by Density Functional Theory (DFT) calculations, and it introduces electron-hole asymmetry in our model.

Interlayer hoppings are given by Slater-Koster-type two-center $pp\pi$ and $pp\sigma$ bond integrals \cite{Slater1954} 
\begin{equation}
    t_{ij} = V_{pp\pi}(1-l^2) + V_{pp\sigma}l^2,
\end{equation}
where $l = \vec{r}_{ij} \cdot \hat{u}_{\perp}/|\vec{r}_{ij}|$ is the cosine of the angle formed between the distance vector $\vec{r}_{ij}$ connecting the two sites $i$ and $j$ and the unit vector $\hat{u}_{\perp}$ perpendicular to the layers. The $\pi$ and $\sigma$ bond integrals are given by 
\begin{gather}
    V_{pp\pi} = t_{\parallel}\exp^{q_{\pi}(1-\frac{|\vec{r}_{ij}|}{a})}\\
    V_{pp\sigma} = t_{\perp}\exp^{q_{\sigma}(1-\frac{|\vec{r}_{ij}|}{d})}.
\end{gather}
$t_{\perp}$ is the $pp\sigma$ integral between two orbitals in opposite layers with same in-plane spatial coordinates and takes the value  $t_{\perp}=0.48$ eV. The decay of bond integrals with increasing interatomic distance is determined by the rates $q_\pi$ and $q_{\sigma}$, which are isotropic and set to $q_\pi/a = q_\sigma/d = 2$. This choice of parameters qualitatively reproduces the band structure of aligned NPG/graphene bilayers in the AB or Bernal stacking sequence as obtained using DFT (see Fig. \ref{FigS1}). Calculation details for the DFT simulations can be found in the next subsection.

For the tight-binding  calculations a $111 \times 416 \times 1$ Monkhorst-Pack \textit{k}-grid \cite{Monkhorst1976} was employed to sample the Brillouin zone of aligned NPG/graphene bilayers ($\theta = 0^\circ$). The number of \textit{k}-points for twisted geometries was scaled according to their lattice parameter size. The tight-binding model and electronic structure calculations were set up using the SISL python package \cite{zerothi_sisl}.

\subsection{Electron transport simulations}

Quantum electron transport simulations were performed using the TRANSIESTA utility TBTRANS \cite{Papior2017}. Combining tight-binding Hamiltonians and Non-Equilibrium Green's Functions (NEGF), TBTRANS allows computing transport in systems up to $\sim 100,000$ atoms. In this work we consider a single-electrode set-up composed of a finite twisted bilayer disk $\sim70$ nm in diameter ($\sim 250,000$ atoms) in point-contact with a metallic tip in the wide-band limit\cite{Calogero2018}. 
The atomic structure of the tip is not explicitly defined, and it is simulated via an on-site energy broadening self-energy on a single NPG site  \cite{Calogero2019b}. For simplicity, in the main text we consider contact to a NPG site in the center of a ribbon and closest to the rotation center of the bilayer. This choice of contact position ensures that for all twist-angles current is injected in a region with local AB stacking  (see Fig. 1 in the main text). In order to avoid backscattering off the finite device edges, we place an isotropic $10$ nm wide complex adsorbing potential (CAP) at the disk boundary \cite{Calogero2018,Calogero2019a,Calogero2019b}, which adsorbs electrons injected from the tip. Although 
in this work 
we restrict ourselves to commensurate twist angles, one of the utilities of applying surrounding CAPs relies on the ability to mimic open boundary conditions and simulate the electronic propagation at any arbitrary (commensurate or incommensurate) twist angle.

Based on the above considerations, the Green's function of the device is given by

\begin{equation}
    \boldsymbol{G} = [\boldsymbol{I}(E+i\eta) - \boldsymbol{H} - i\boldsymbol{V_c} - i\boldsymbol{\Gamma_t}]^{-1},
\end{equation}

where $\boldsymbol{H}$ is the Hamiltonian matrix of the twisted bilayer disk obtained from Eq.\,\ref{Eq1}, $i\boldsymbol{V_c}$ is the diagonal CAP matrix, and $i\boldsymbol{\Gamma_t}$ is the on-site self-energy of the tip in the wide-band limit, with $\boldsymbol{\Gamma_t}$ the decay rate into the tip. The flow of electrons injected from the tip into the device can then be visualized by computing bond-transmissions from the Green's function (see Fig. 1 in the main text) \cite{Calogero2019a,Calogero2019b}.

\newpage

\section{Additional electron transport simulations}

\begin{figure*}[h]
\begin{center}
\includegraphics[width=\linewidth]{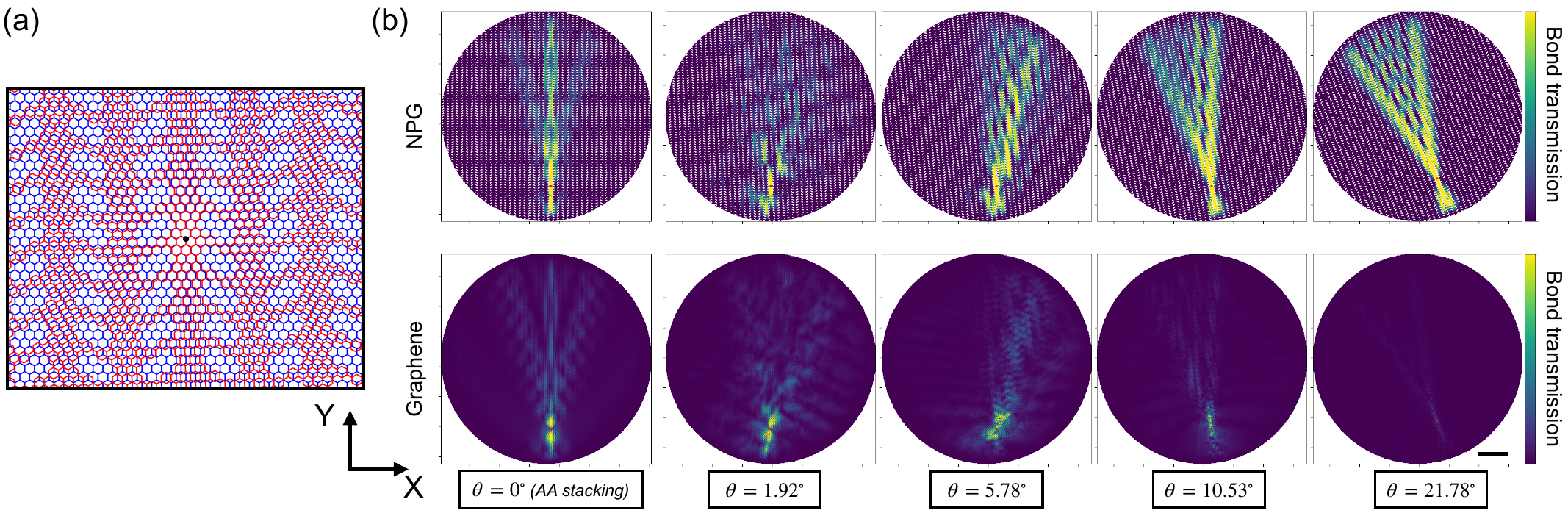}
\caption{(a) Atomic structure of a twisted NPG/graphene bilayer
in an AA stacking configuration, 
where the black dot indicates the
selected rotation center.
(b) Real-space bond transmissions at an energy $E=-0.4$ eV for different twist angles of the NPG/graphene bilayer shown in (a). The red dot indicates the injection point in the NPG layer, which coincides with the rotation center (black dot in panel (a)). Scale bar is 10 nm. 
}
\label{FigS2}
\end{center}
\end{figure*}

In order to test the robustness of our conclusions against structural details, such as initial stacking, rotation axis or injection position, here we present additional transport simulations. We calculate real-space bond transmissions in NPG/graphene bilayers initially in AA stacking ($\theta = 0^\circ$), and we rotate the NPG layer with respect to an axis crossing atoms of both layers in an AA site, as depicted in Fig. \ref{FigS2}a. As shown in Fig. \ref{FigS2}b, the corresponding in-plane bond transmissions follow the same
trend as discussed in the main text, exhibiting gradual interlayer electronic decoupling and transport asymmetries. 

\section{Reciprocal space analysis}

\begin{figure*}[t]
\begin{center}
\includegraphics[width=5in]{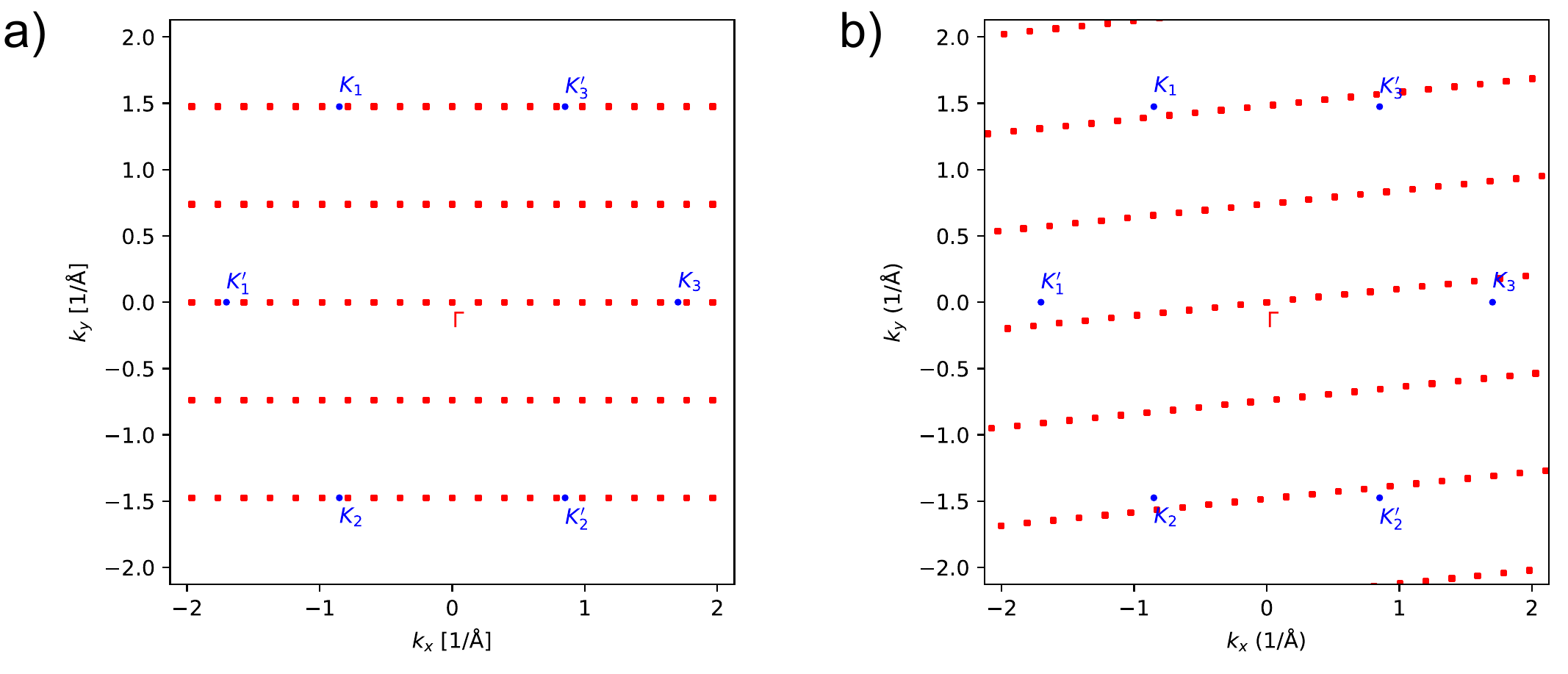}
\caption{Superimposed reciprocal lattices of graphene and NPG. The reciprocal lattice of NPG appears rotated by $\theta=0^\circ$ (a) and $\theta=5.78^\circ$ (b). $K/K'$ points of graphene and $\Gamma$-points of  NPG are indicated by blue circles and red squares, respectively.
}
\label{FigS3}
\end{center}
\end{figure*}

In the aligned configuration (AA or AB stacking) the NPG/graphene bilayer has mirror symmetry defined by a plane
parallel to the growth
direction of the GNRs and crossing the center of its backbone. The mirror symmetry is broken upon twisting, and only time-reversal symmetry is retained. The latter relates $K_1 \rightarrow K'_2$, $K'_1 \rightarrow K_3$ and $K_2 \rightarrow K'_3$ as shown in
Fig.\,\ref{FigS3}. Therefore it is sufficient to focus on $K'_2$, $K_3$ and $K'_3$ to analyze the overlap in momentum-space between constant energy contours of NPG and graphene.

\begin{figure*}[h!]
\begin{center}
\includegraphics[width=\linewidth]{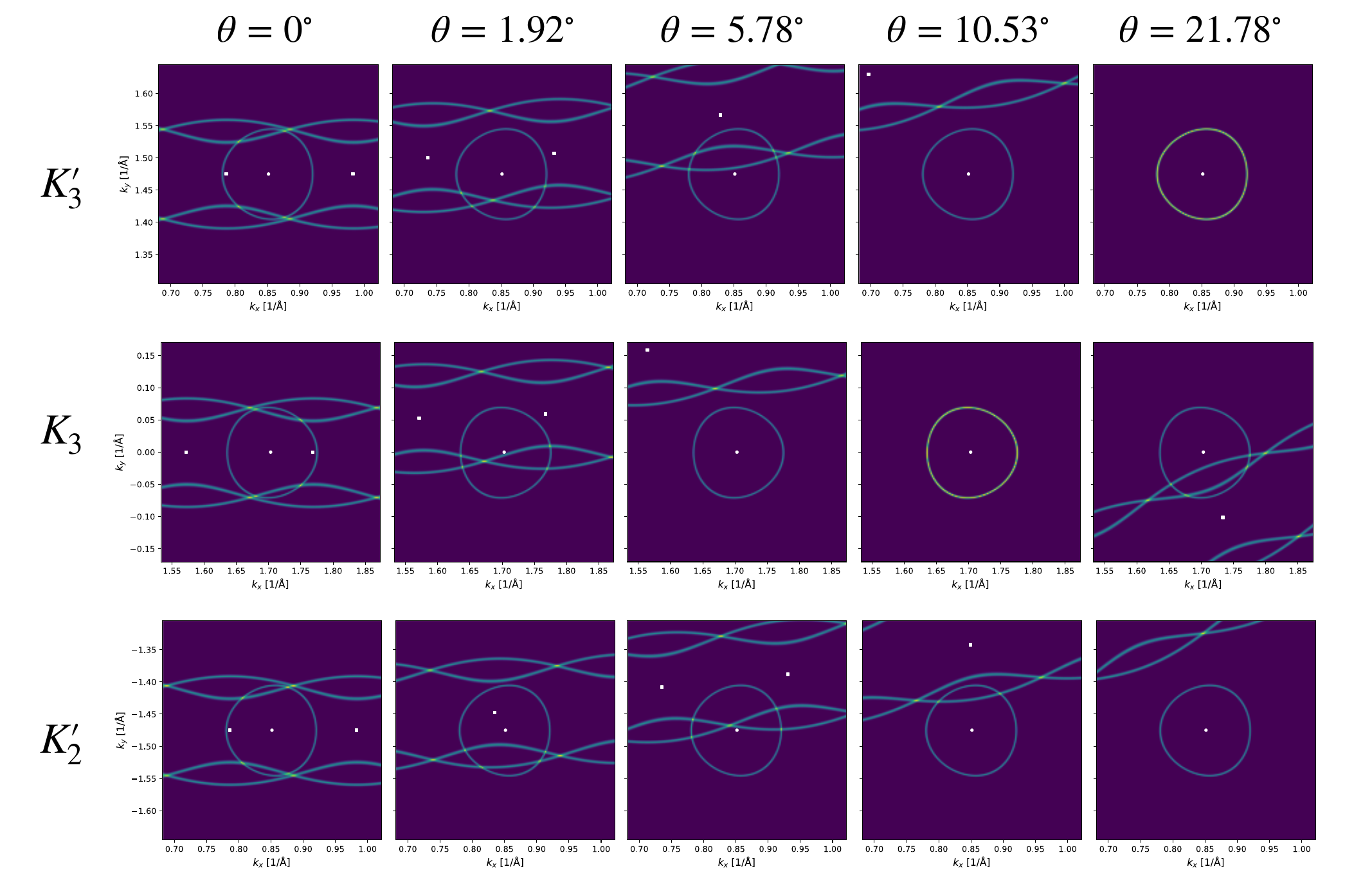}
\caption{Superimposed constant energy contours for twisted NPG/graphene at E=-0.4 eV, around the three inequivalent $K$-points of graphene in a generalized momentum space before folding.
}
\label{FigS4}
\end{center}
\end{figure*}

Fig. \ref{FigS4} shows constant energy contours of NPG and graphene at an energy $E=-0.4$ eV in a generalized momentum space before folding. 
Although the $K/K'$ points shown are inequivalent, at small $\theta$, the general trend as $\theta$ increases is an increasing separation of NPG bands from graphene Dirac cones, which suppresses the interlayer coupling. The panels corresponding to $\theta = 21.78^\circ$ show a change of trend, with the NPG bands approaching the Dirac cones around $K_3$. In this case, the interlayer coupling is quenched by wavefunction symmetry.

\newpage

\newpage

\section{Electronic structure origin of chiral currents}

\begin{figure*}[h]
\begin{center}
\includegraphics[width=\linewidth]{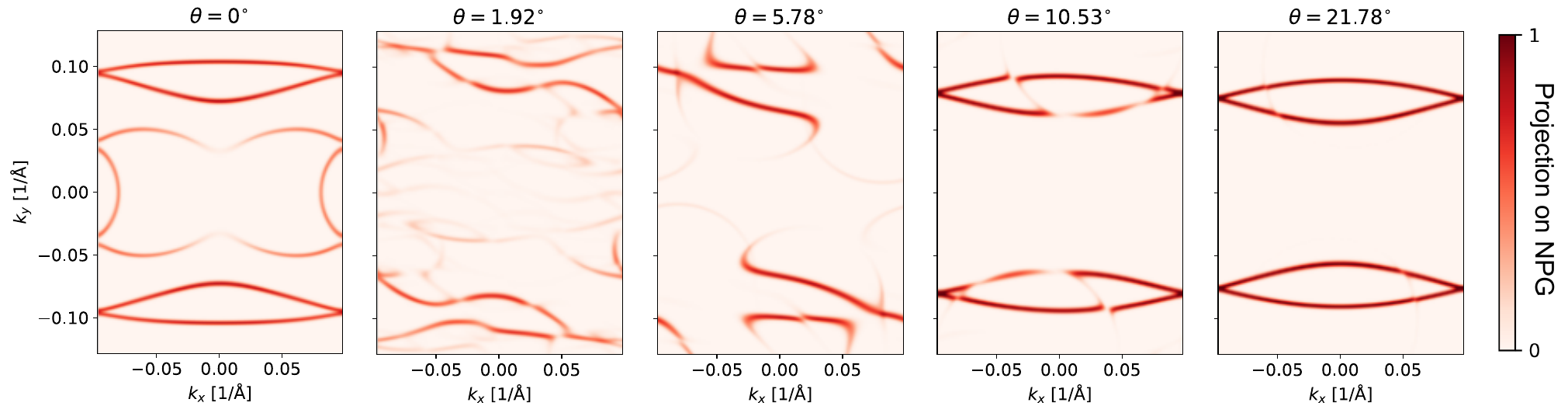}
\caption{Constant energy contours of NPG/graphene at $E=-0.4$ eV and different $\theta$, projected onto the electronic states of NPG and unfolded to its Brillouin zone.
}
\label{FigS5}
\end{center}
\end{figure*}

Similarly to Fig. 2 in the main text, Fig. \ref{FigS5} shows constant energy contours of the band structure of NPG/graphene bilayers at $E=-0.4$ eV, projected onto the electronic states of NPG and unfolded to its Brillouin zone. The unfolding procedure 
allows us to rationalize some of the features exhibited by the in-plane electronic propagation within NPG. 

At AB stacking ($\theta = 0^\circ$) the energy contour is characterized by the mirror symmetry of the system, with symmetric band dispersion with respect to the Y-axis (i.e. the axis parallel to the GNR growth direction). As a combination of the interlayer coupling and the twist-induced mirror-symmetry-breaking, constant energy contours of NPG appear strongly and asymmetrically (with respect to the Y-axis) modified at small $\theta$. As $\theta$ keeps increasing, energy contours of monolayer NPG are gradually restored, with avoided crossings unevenly distributed with respect to the Y-axis (see $\theta = 10.53^\circ$). This asymmetric interaction of NPG bands with graphene Dirac cones breaks the symmetry of the dispersion relation in the direction across GNRs and leads to the asymmetric propagation shown in Fig. 1 in the main text. For large enough twist-angles, the interlayer coupling becomes so weak that the mirror-symmetry-breaking becomes irrelevant and monolayer transport behaviour is restored (see $\theta = 21.78^\circ$).

\section{Additional LDOS map simulations}

\begin{figure*}[h]
\begin{center}
\includegraphics[width=2.5in]{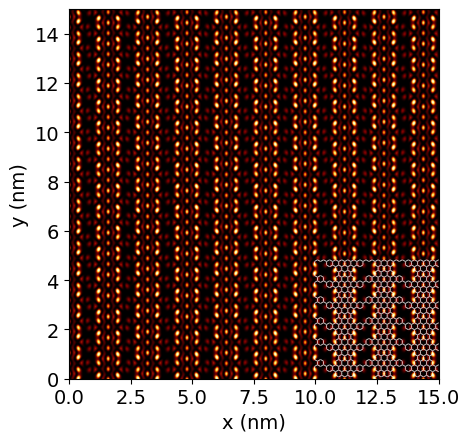}
\caption{LDOS map of monolayer NPG at the onset energy of the VB ($E = -195$ meV). The NPG geometry is overlaid in the bottom right corner of the figure.
}
\label{FigS6}
\end{center}
\end{figure*}

\bibliography{bibliography}